\documentclass[fleqn,twoside]{article}
\usepackage[headings]{espcrc2}
\readRCS
$Id: espcrc2.tex,v 1.2 2004/02/24 11:22:11 spepping Exp $
\usepackage{graphicx,cite}
\usepackage[figuresright]{rotating}
\def\sing{\hbox{\tiny sing.}}
\def\reg{\hbox{\tiny reg.}}
\def\tot{\hbox{\tiny tot}}
\def\MSbar{\hbox{\tiny ${\overline{\rm MS}}$}}
\catcode`\@=11 
\def\lsim{\mathrel{\mathpalette\@versim<}}
\def\gsim{\mathrel{\mathpalette\@versim>}}
\def\@versim#1#2{\vcenter{\offinterlineskip
        \ialign{$\m@th#1\hfil##\hfil$\crcr#2\crcr\sim\crcr } }}

\newcommand{\AmS}{{\protect\the\textfont2
  A\kern-.1667em\lower.5ex\hbox{M}\kern-.125emS}}

\title{Progress in computing inclusive B decay spectra }

\author{Einan Gardi and Jeppe R. Andersen
\\
\vspace{1pc} {Cavendish Laboratory, University of Cambridge, \\J J
Thomson Avenue, Cambridge, CB3 0HE, UK}
}

\runtitle{} \runauthor{E. Gardi and J.R. Andersen}

\begin{document}

\begin{abstract}
We review the progress in the QCD calculation of inclusive decay
spectra. It was recently shown that the inherent infrared finiteness
of inclusive spectra extends beyond the level of logarithms. Dressed
Gluon Exponentiation makes practical use of this property by
computing the Sudakov exponent as a Borel sum. Based on renormalon
analysis, infrared sensitivity in the exponent is reflected in power
corrections that are inversely proportional to the third power of
the mass. Therefore, the parametric enhancement of non-perturbative
corrections near the phase--space boundary is effective only in a
small region. Consequently, the on-shell decay spectrum provides a
good approximation to the meson decay. In particular, it facilitates
a precise determination of $|V_{ub}|$ from present measurements of
inclusive charmless semileptonic widths without involving a
non-perturbative ``shape function''.
\end{abstract}

\maketitle


One of the important challenges for QCD presented by the B factories
is the calculation of inclusive partial decay widths within specific
regions of phase space. The obvious example is the inclusive
measurement of charmless semileptonic decay, ${\bar
B}\longrightarrow X_ul\bar{\nu}$, used for the determination
of~$|V_{ub}|$. Owing to the overwhelming charm background, this
measurement is strictly restricted to the region of small hadronic
mass, $M_X< 1.7$ GeV, where charmed final states are kinematically
excluded; specific cuts within this region are adopted depending on
the experimental techniques applied. Consequently, extracting
$|V_{ub}|$ relies on quantitative theoretical understanding of the
spectrum.

The common lore has been that estimates of the partial widths for
the relevant cuts strongly depend on the non-perturbative structure
of the meson. The spectrum in the small--$M_X$ region was
obtained~\cite{Falk:1997gj,Bigi:1997dn,Lange:2005yw} through the
convolution of a computable hard coefficient function with a
leading--twist momentum distribution function of the b quark in the
meson, the ``shape function''~\cite{Neubert:1993um,Bigi:1993ex}.
Being non-perturbative, the latter was not computed, but rather
parametrized and fitted to the measured ${\bar B}\longrightarrow X_s
\gamma$ spectrum, in analogy with deep inelastic (DIS) structure
function phenomenology. The ``shape function'' has been the dominant
source of \hbox{uncertainty~in determining~$|V_{ub}|$}.

Recently the situation has changed. It was shown that the inherent
infrared finiteness of inclusive spectra extends beyond the level of
logarithms~\cite{BDK}: the leading renormalon ambiguity in the
Sudakov exponent cancels against that of the pole mass while the
next--to--leading one is absent. Upon using Dressed Gluon
Exponentiation~(DGE)~\cite{DGE_thrust,Gardi:2002bg,Gardi:2001di,CG,Gardi:2003ar,GR}
the ${\bar B}\longrightarrow X_s \gamma$ and ${\bar
B}\longrightarrow X_ul\bar{\nu}$ decay spectra can be well
approximated by the corresponding on-shell decay
spectra~\cite{BDK,Andersen:2005bj,QD,Andersen:2005mj,Gardi:2005mf,Gardi:2004gj}.
Non-perturbative effects associated with the meson structure enter
in this framework as power corrections, and have just a small impact
on the experimentally--relevant partial widths. The purpose of this
talk is to explain why this is so.



Inclusive B decays such as $\bar{B}\longrightarrow X_s \gamma$ and
$\bar{B}\longrightarrow X_u l \bar{\nu}$ are dominated by jet-like
momentum configurations: the hadronic system $X$ has a small mass,
$M_X\ll M_B$, while its energy in the B rest frame is large.
Inclusive decays can be analyzed within perturbation theory owing to
the fact that the heavy quark carries most of the meson momentum and
it is therefore close to its mass shell. By making the perturbative,
on-shell approximation one neglects non-perturbative effects that
are suppressed by inverse powers of the heavy--quark mass. In the
region of interest, of small $M_X$, these power corrections are
parametrically enhanced. This will be a central issue in what
follows.

Given the typical jet--like momentum configuration it is convenient
to compute the spectrum in terms of lightcone coordinates. The
hadronic lightcone coordinates are defined by: $P^{\pm}=E_X \mp
|\vec{P}_X|$. The jet is characterized\footnote{In
$\bar{B}\longrightarrow X_s \gamma$ the photon is real so $P^-\equiv
M_B$, while in $\bar{B}\longrightarrow X_u l \bar{\nu}$, where the
lepton pair has a mass $q^2=(M_B-P^+)(M_B-P^-)$, the distribution in
$P^-$ is rather broad but still peaks near $P^- \sim  M_B$, see
Fig.~\ref{fig:diff_Pminus_different_cuts}.} by large $P^-$, of
${\cal O} (M_B)$ and small $P^+$, which is of the order of the mass
difference between the meson and the quark\footnote{Here and below
$m_b$ stands for the quark pole mass. The renormalon
ambiguity~\cite{Beneke:1994sw,Bigi:1994em} is dealt with explicitly
using the Principal Value prescription.}, $\bar{\Lambda}=M_B-m_b$.
Within the on-shell approximation, we define the partonic lightcone
coordinates: $p_j^{\pm}=P^{\pm}-\bar{\Lambda}$. At Born level, the
$p_j^+$ distribution is trivial, $\delta(p_j^+)$. Beyond this order,
the small $p_j^+$ region is characterized by soft and collinear
gluon emission that broadens the $p_j^+$ distribution. This is why
Sudakov resummation is essential in computing the spectrum.

One obvious limitation of perturbation theory is the fact that the
perturbative spectrum, at any order, has support only for $p_j^+>0$,
i.e. for \hbox{$P^+>\bar{\Lambda}$}, while the physical spectrum
fills up the whole region down to $P^+=0$. As we shall see below,
this inherent limitation is removed in the DGE
approach~\cite{Andersen:2005bj}.

In perturbation theory, the large hierarchy between $p_j^+$ and
$p_j^-$ in the peak region is reflected in large Sudakov logarithms.
It is convenient to choose the resummation variable as the exponent
of the rapidity, $r\equiv p_j^+/p_j^-$ and write the triple
differential width in $b\longrightarrow X_u l \bar{\nu}$
as~\cite{Andersen:2005mj}
\begin{eqnarray}
\label{triple_diff_r}
\frac{1}{\Gamma_0}\frac{d\Gamma(\lambda,r,x_l)}{d\lambda d r dx_l}
=V(\lambda,x_l) \delta(r)+ R(\lambda,r,x_l),
\end{eqnarray}
where the additional kinamatic variables are $\lambda=p_j^-/m_b$ and
$x_l=2E_l/m_b$, the Born--level width is
$\Gamma_0={G_F^2\left|V_{ub}\right|^2m_b^5}/{(192 \pi^3)}$, the
virtual corrections are
\begin{eqnarray}
\label{V_and_D_r} &&\hspace*{-20pt}V(\lambda,x_l)=w_0(\lambda,x_l)
+\frac{C_F\alpha_s}{\pi}w_1(\lambda,x_l)+\cdots
\end{eqnarray}
and the real--emission contribution, which starts at~${\cal
O}(\alpha_s)$, is further split as
$R(\lambda,r,x_l)=R_{\sing}(\lambda,r,x_l)+R_{\reg}(\lambda,r,x_l)$.
The regular piece $R_{\reg}(\lambda,r,x_l)$, similarly to
$V(\lambda,x_l)$, is known~\cite{DFN,Andersen:2005mj} to ${\cal
O}(\alpha_s)$  only, while\footnote{For the definition of the $()_*$
distribution and the details of the NLO result in these variables
see Sec.~3.1~in~Ref.~\cite{Andersen:2005mj}.}
\[
\!R_{\sing}(\lambda,r,x_l)\!=\!-w_0(\lambda,x_l)\!\left(\frac{\ln
r}{r}\! +\!
\frac{7}{4}\frac{1}{r}\right)_{*}\!\!\!\frac{C_F\alpha_s}{\pi}\!+\!\cdots
\]
which contains \emph{all the non-integrable terms} at
$r\longrightarrow 0$, regularized as $()_*$ distributions, is known
in full to ${\cal O}(\alpha_s^2)$ --- see Eq. (3.41) in
Ref.~\cite{Andersen:2005mj}
---  and in part\footnote{The~\emph{exponent} in Eq.~(\ref{alt_Sud_SL}) below is known to NNLL
accuracy. Nevertheless, in order to determine
$R_{\sing}(\lambda,r,x_l)$ to this accuracy at ${\cal
O}(\alpha_s^3)$ and beyond one would need to compute the NNLO
virtual coefficient $w_2(\lambda,x_l)$. }, at higher orders. This
higher--order information is contained in the resummation formula:
\begin{eqnarray}
\label{r_mom_def} \frac{d\Gamma_{N} \nonumber
 (\lambda,x_l)}{d\lambda dx_l}
  \!\!\!\!&\equiv&\!\!\!\!\frac{1}{\Gamma_0}\int_{0}^{(1-x_l)/\lambda} \!\!\!\!\!\!\!\!\!\!\!\!\!\!
 dr
 \left(1-r\right)^{N-1}
\frac{d\Gamma(\lambda,r,x_l)}{d\lambda dr dx_l}
 \\
&&\hspace*{-55pt}= V(\lambda,x_l)\times  {\rm
Sud}(m_b\lambda,N)+\Delta R_N(\lambda,x_l).
\end{eqnarray}
In the small $p_j^+$ region, corresponding to large $N$, the first
term dominates: it sum up to all orders perturbative corrections
that diverge as powers of $\ln N$ or are finite in the
$N\longrightarrow \infty$ limit. The remainder,
\begin{eqnarray}
&&\hspace*{-20pt}\Delta R_N(\lambda,x_l) \equiv
\int_0^{{(1-x_l)}/{\lambda}} \!\!\!\!\!\!\!\!\!dr (1-r)^{N-1}
R_{\reg}(r,\lambda,x_l),\nonumber
\end{eqnarray}
is of~${\cal O}(1/N)$ and can be included at fixed order.


The resummation formula (\ref{r_mom_def}) is a manifestation of the
\emph{infrared and collinear safety} of the on-shell decay spectrum:
owing to the cancellation of infrared singularities between real and
virtual corrections, the moments have finite expansion coefficients
to any order in perturbation theory. As usual, this cancelation
leaves behind large Sudakov logarithms. These arise from two
distinct
subprocesses~\cite{Korchemsky:1994jb,Bauer:2000yr,BDK,Bosch:2004th}:
the formation of a jet of invariant mass squared of ${\cal
O}(p^+_jp^-_j)$ and radiation off the nearly on-shell b--quark,
characterized by transverse momenta of ${\cal O}(p^+_j)$. The two
corresponding functions, the \emph{jet function}
$J_N$~\cite{Sterman:1986aj,CT,Contopanagos:1996nh,GR} and the
\emph{quark distribution in an on-shell heavy
quark}~$S_N$~\cite{QD,Korchemsky:1992xv,KR87_92} obey the following
evolution equations in moment space:
\begin{eqnarray}
\label{JS_evolution} &&\hspace*{-21pt}\frac{d\ln J_N(Q;\mu_F)}{d\ln
Q^2}=
\!\!\int_0^1 \!\!\frac{dr}{r}\!\!\left(\bar{r}^{N-1}-1\right) {\cal
J}\left(\alpha_s(rQ^2)\right),\nonumber\\ \\
&&\hspace*{-21pt} \frac{d\ln S_N(Q;\mu_F)}{d\ln
Q^2}=
\!\!-\!\!\int_0^1 \!\!\frac{dr}{r}\!\! \left(\bar{r}^{N-1}-1\right)
{\cal S}\left(\alpha_s(r^2Q^2)\right), \nonumber
\end{eqnarray}
where  $\bar{r}\equiv 1-r$ and where ${\cal J}$ and~${\cal S}$ are
Sudakov anomalous dimensions that are now known in full to
NNLO~\cite{MVV,Andersen:2005bj,QD,Korchemsky:1992xv}. The Sudakov
factor, summing up the logarithmically--enhanced terms in decay
spectra to all orders, is: ${\rm Sud}(\lambda m_b,N)=J_N(\lambda
m_b;\mu_F)S_N(\lambda m_b;\mu_F)$, where the factorization--scale
dependence cancels exactly in the product.

Note the fundamental difference with DIS structure functions where
there is just one source~\cite{GR} of Sudakov logarithms, namely the
jet function $J_N(Q;\mu_F)$, and the factorization--scale dependence
cancels against the non-perturbative quark distribution function
$q_N(\mu_F)$, e.g. $F_2^N \sim J_N(Q;\mu_F) q_N(\mu_F)$. Contrary to
$S_N$, $q_N$ cannot be defined in perturbation theory owing to the
collinear singularity of an incoming light quark.

Because of the cancellation of \emph{logarithmic} infrared
singularities, the evolution kernels defined by the
r.h.s.~in~(\ref{JS_evolution}) are finite to any order in
perturbative theory. However, these kernels conceal infrared
sensitivity at the \emph{power}
level~\cite{Contopanagos:1993yq,Grozin:1994ni,Korchemsky:1994is,DGE_thrust,Gardi:2002bg,Gardi:2001di,CG,Gardi:2003ar,GR,BDK,Andersen:2005bj,QD,Andersen:2005mj,Gardi:2005mf,Gardi:2004gj},
which only becomes explicit once \emph{running--coupling effects}
are resummed to all
orders~\cite{DGE_thrust,Gardi:2002bg,Gardi:2001di,CG,Gardi:2003ar,GR}.
A systematic way to quantify this infrared sensitivity is to
regularize the ensuing divergence of the perturbative expansion by
Borel summation~\cite{Mueller:1984vh,Beneke:1998ui,Beneke:1995pq}.
To this end one writes the scheme--invariant Borel
representation~\cite{Grunberg:1992hf} of the anomalous
dimensions~\cite{DGE_thrust,Gardi:2002bg,Gardi:2001di,CG,Gardi:2003ar,GR,BDK,Andersen:2005bj,QD,Andersen:2005mj,Gardi:2005mf,Gardi:2004gj}:
\begin{eqnarray}
\label{Borel_rep_JS} &&\hspace*{-20pt} {\cal
J}\left(\alpha_s(\mu^2)\right)\!=\!\frac{C_F}{\beta_0}
\int_0^{\infty}\!du
\left(\frac{\Lambda^2}{\mu^2}\right)^u T(u) B_{\cal J}(u), \\
\nonumber &&\hspace*{-20pt} {\cal
S}\left(\alpha_s(\mu^2)\right)\!=\!\frac{C_F}{\beta_0}
\int_0^{\infty}\!du \left(\frac{\Lambda^2}{\mu^2}\right)^u T(u)
B_{\cal S}(u),
\end{eqnarray}
where $\beta_0=\frac{11}{12}C_A-\frac{1}{6}N_f$ and $T(u)\equiv
(u\delta)^{u\delta}{\rm e}^{-u\delta}/\Gamma(1+u\delta)$ with
$\delta\equiv \beta_1/\beta_0^2$. Using (\ref{Borel_rep_JS}) in
(\ref{JS_evolution}) one can explicitly perform the $r$ integration
and identify potential power--like ambiguities arising from the
$r\longrightarrow 0$ limit as Borel singularities. The solution of
the evolution equations, formulated as a Borel sum,
is~\cite{Andersen:2005bj}:
\begin{eqnarray}
\label{alt_Sud_SL} &&\hspace*{-20pt}{\rm Sud}(\lambda m_b,N)
=\exp\bigg\{ \frac{C_F}{\beta_0}\int_0^{\infty}\frac{du}{u} \, T(u)
\left(\frac{\Lambda^2}{\lambda^2m_b^2}\right)^u
 \nonumber \\  && \hspace*{-10pt}
\times \bigg[ B_{\cal
S}(u)\left(\frac{\Gamma(-2u)\Gamma(N)}{\Gamma(N-2u)}+\frac{1}{2u}\right)
\\ \nonumber &&\hspace*{30pt}
 - B_{\cal
J}(u)\left(\frac{\Gamma(-u)\Gamma(N)}{\Gamma(N-u)}+\frac{1}{u}\right)
\bigg] \bigg\}.
\end{eqnarray}
In the jet--function part one finds potential renormalon ambiguities
at positive integer values of~$u$, while in the soft--function part
at any integer and half integer~$u$. Being anomalous dimensions,
$B_{\cal J}(u)$ and $B_{\cal S}(u)$  are not expected to have any
renormalon singularities of their own, however, unless these
functions \emph{vanish} at these locations there will be power-like
ambiguities in the evolution kernels in~(\ref{JS_evolution}). These
 ambiguities scale as powers of $\Lambda^2 N /(\lambda m_b)^2$ and
$\Lambda N/(\lambda m_b)$, corresponding to the jet--mass and the
soft scale, respectively, so they are parametrically--enhanced at
large~$N$.

The discussion of renormalon singularities can be made concrete by
focusing on the gauge--invariant set of radiative corrections
corresponding to the large--$\beta_0$ limit\footnote{Results in this
limit are obtained~\cite{Beneke:1998ui} by first considering the
large--$N_f$ limit, in which a gluon is dressed by any number of
fermion--loop insertions, and then making the formal substitution
$N_f\longrightarrow -6 \beta_0$.}. The anomalous dimensions are then
obtained as analytic functions in the Borel plane~\cite{GR}:
\begin{eqnarray}
\label{B_DJ_large_beta0} \left.B_{\cal J}(u)\right\vert_{{\rm large}
\, \beta_0}&\!\!\!\!=\!\!\!\!&\frac{{\rm e}^{\frac53 u}}{2}
\left(\frac{1}{1-u}+\frac{1}{1-u/2}\right)
\frac{\sin\pi u}{\pi u}
,\nonumber \\
\left.B_{\cal S}(u)\right\vert_{{\rm large} \,
\beta_0}&\!\!\!\!=\!\!\!\!&{\rm e}^{\frac53 u}(1-u)
.
\end{eqnarray}
Based on these expressions one can deduce which power ambiguities
indeed appear in Eq.~(\ref{alt_Sud_SL}). On the soft scale one finds
ambiguities corresponding to \hbox{ $u=\frac12$} and $u\geq \frac32$
and on the jet mass scale, $u=1$ and $u=2$. Of course, these
ambiguities should all cancel once non-perturbative corrections are
systematically included. Conversely, in absence of a
non-perturbative calculation, the ambiguities provide a good hint on
the functional form, and quite possibly even the magnitude, of the
non-perturbative power corrections.

Although the dominance of non-perturbative corrections that are
probed by renormalons cannot be established from first principles in
QCD, this conjecture has led to successful power--correction
phenomenology in a variety of
applications~\cite{Dokshitzer:1995qm,Beneke:1998ui,Dasgupta:2003iq,Gardi:1999dq}.
The DGE approach was applied to several infrared and collinear safe
observable, notably event--shape
distributions~\cite{DGE_thrust,Gardi:2002bg} and heavy--quark
fragmentation~\cite{CG,Gardi:2003ar} where precise data from LEP was
used to test the renormalon dominance assumption. The observed
hadronization effects near the phase--space boundary were found to
be in agreement with the pattern of power corrections predicted by
renormalons. On these grounds one expects that the renormalon
structure of the Sudakov exponent provides a good indication on
non-perturbative corrections also in B decay spectra. Of course,
here the leading corrections are associated with the initial--state
B meson, rather than final--state hadronization, and the renormalon
dominance assumption must be again confronted with data.


In computing the Sudakov factor (\ref{alt_Sud_SL}) we assume that
the pattern of renormalon singularities of the large--$\beta_0$
limit holds in the full theory. Nonetheless, the perturbative
expansions of these functions at $u=0$ and likewise the residues in
(\ref{alt_Sud_SL}) get modified by ${\cal O}(1/\beta_0)$
contributions. In order to match (\ref{B_DJ_large_beta0}) onto the
perturbative expansions of ${\cal J}$ and ${\cal S}$~\cite{QD}, we
write the following ansatz~\cite{Andersen:2005bj,Andersen:2005mj}:
\begin{eqnarray}
\label{B_JS_QCD} B_{\cal
J}(u)&\!\!\!\!\!\!=\!\!\!\!\!\!&\left.B_{\cal J}(u)\right\vert_{{\rm
large} \, \beta_0}\times {\rm e}^{c_1u +c^{\cal J}_2u^2} \times
W_{\cal J}(u)\\\nonumber B_{\cal
S}(u)&\!\!\!\!\!\!=\!\!\!\!\!\!&\left.B_{\cal S}(u)\right\vert_{{\rm
large} \, \beta_0}\times {\rm e}^{c_1u +c^{\cal S}_2u^2} \times
W_{\cal S}(u),
\end{eqnarray}
where~$c_1=\left(1-{\pi^2}/{4}\right){C_A}/({3\beta_0})$ is
universal~\cite{KR87_92}, $c^{{\cal J}/{\cal S}}_2$ are determined
based on the known NNLO expansions of these functions and $W_{{{\cal
J}/{\cal S}}}(u)$ parametrize yet--unknown ${\cal O}(u^3)$
corrections. By fixing $B_{\cal S}(u)$ and $B_{\cal J}(u)$ this way
and choosing the Principal Value prescription for the renormalons,
the Sudakov factor (\ref{alt_Sud_SL}) is uniquely determined.

Now, the resummed spectrum is obtained by an inverse Mellin
transformation of Eq.~(\ref{r_mom_def}),
\begin{eqnarray}
\label{sl_logs_resummed_r}  \frac{1}{\Gamma_0}
\frac{d\Gamma(\lambda,r,x_l)}{d\lambda dr dx_l }=
 \int_{\cal C}\frac{d N}{2\pi i}
 \left(1-{r}\right)^{-N}\!
\frac{d\Gamma_{N}
 (\lambda,x_l)}{d\lambda dx_l},
\end{eqnarray}
where the integration contour ${\cal C}$ runs parallel to the
imaginary axis, to the right of the singularities of the integrand.
Finally, the triple differential distribution in physical, hadronic
variables is:
\begin{eqnarray}
\label{Pplus_diff} &&\hspace*{-15pt} \frac{1}{\Gamma_0}
\frac{d\Gamma(P^+,\,P^-,\,E_l)}{dP^+\, dP^-\,dE_l} =\frac{2}{\lambda
m_b^3}\,\times
\\\nonumber &&\hspace*{-15pt}\left.\frac{1}{\Gamma_0}
\frac{d\Gamma(\lambda,r,x_l)}{d\lambda dr dx_l
}\right\vert_{\left\{r= \frac{P^+ -
\bar{\Lambda}}{P^--\bar{\Lambda}};\,
\lambda=\frac{P^--\bar{\Lambda}}{m_b};\,
x_l=\frac{2E_l}{m_b}\right\} }.
\end{eqnarray}

\begin{figure}[t]
\includegraphics[angle=90,width=18pc]{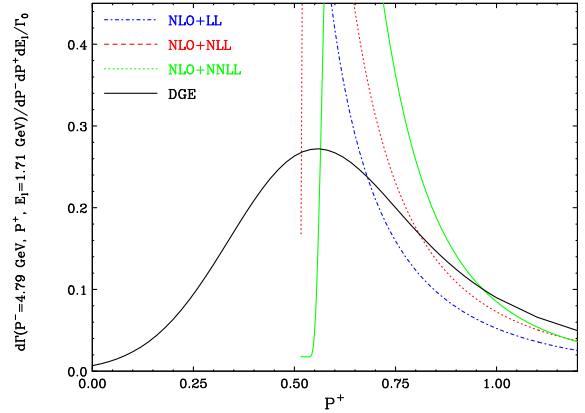}
\vspace*{-30pt} \caption{The $P^+$ spectrum in
$\bar{B}\longrightarrow X_u l \bar{\nu}$ in the on-shell
approximation at a representative point in phase space: $P^-=4.79$
GeV; $E_l=1.71$ GeV. The spectrum is computed with fixed logarithmic
accuracy (LL, NLL, NNLL) and by DGE. All results are matched to
NLO~\cite{Andersen:2005mj}.\label{fig:FLA_vs_DGE}} \vspace*{-10pt}
\end{figure}

The change of variables in (\ref{Pplus_diff}) has a subtle but
absolutely crucial role in obtaining the correct distribution within
the on-shell approximation. To understand it observe that this
transformation involves $\bar{\Lambda}\equiv M_B-m_b$ that is
ambiguous since the pole mass $m_b$ has a $u=\frac12$
renormalon~\cite{Beneke:1994sw,Bigi:1994em}, and recall that the
Sudakov exponent in (\ref{alt_Sud_SL}) also has a $u=\frac12$
renormalon ambiguity. These ambiguities have the same source namely
the ambiguous definition of an on-shell state and they \emph{cancel
out exactly}~\cite{BDK} in Eq.~(\ref{Pplus_diff}). This cancellation
was checked explicitly in the large--$\beta_0$ limit and it is
understood to be general. In the course of the calculation we deal
with it by using the Principal Value Borel sum regularization in
both Eq.~(\ref{alt_Sud_SL}) and in the calculation of the pole
mass~\cite{Andersen:2005bj} in $\bar{\Lambda}$.

\begin{figure}[t]
\includegraphics[angle=0,width=19.5pc]{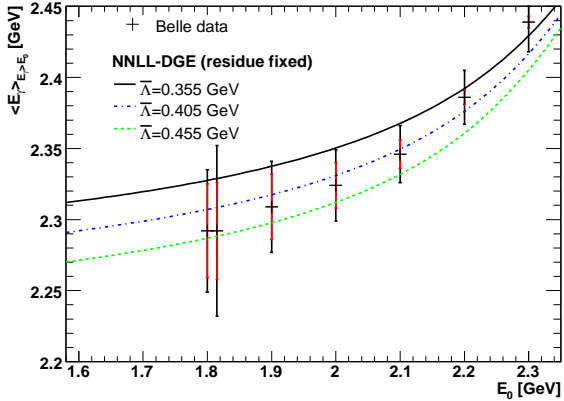}
\vspace*{-30pt} \caption{\label{cut_mom_av}The average energy in
$\bar{B}\longrightarrow X_s\gamma$, $\left<E_{\gamma}\right>$, in
the on-shell approximation,~as~a function of the minimum photon
energy cut $E_0$, as calculated by DGE in~\cite{Andersen:2005bj}
varying $m_b^{\MSbar}$ within its error range. The result is
compared with data from Belle~\cite{Belle05}. Inner and total error
bars show systematic and statistical plus systematic errors (added
in quadrature), respectively.} \vspace*{-10pt}
\end{figure}
\begin{figure}[t]
\includegraphics[angle=0,width=19.5pc]{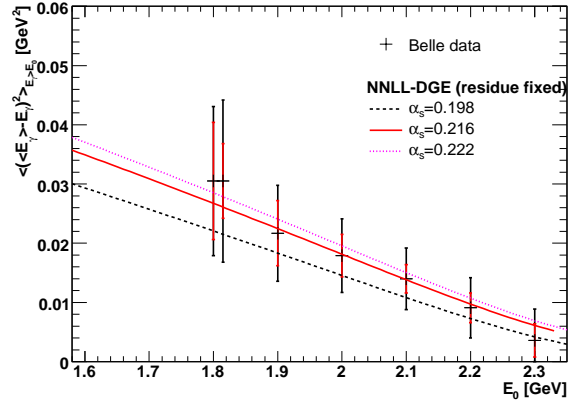}
\vspace*{-30pt} \caption{\label{cut_mom_va}The energy variance in
$\bar{B}\longrightarrow X_s\gamma$,
$\left<(\left<E_{\gamma}\right>-E_{\gamma})^2\right>$,~as~a function
of the minimum photon energy cut $E_0$, as calculated by DGE
in~\cite{Andersen:2005bj} varying $\alpha_s$ within its error range,
compared with data from Belle~\cite{Belle05}. } \vspace*{-10pt}
\end{figure}

The extent to which the perturbative on-shell spectrum is
constrained depends primarily on having the soft anomalous dimension
function $B_{\cal S}(u)$ entering Eq.~(\ref{alt_Sud_SL})
sufficiently well under control. Eq.~(\ref{B_JS_QCD}) incorporates
the QCD perturbative expansion of $B_{\cal S}(u)$ around the origin
up to~NNLO.  However, owing to the soft scales probed as $N$ gets
large, which are ${\cal O}(\lambda m_b/N)$, the Borel sum has some
sensitivity to $B_{\cal S}(u)$ away from the origin. This
sensitivity depends on what is assumed about $W_{\cal S}(u)$ in
Eq.~(\ref{B_JS_QCD}). Here one can make use~\cite{Andersen:2005bj}
of additional information, namely the value of the leading
renormalon residue of the pole mass\footnote{The pole--mass
$u=\frac12$ renormalon residue has been accurately determined,
see~Refs.~\cite{Pineda:2001zq,Lee:2002px,Andersen:2005bj}.}, which
fixes $B_{\cal S}(u=\frac12)$ owing to the exact cancellation of
ambiguities explained above. We further assume that $W_{\cal S}(u)$
has no Borel singularities and thus the vanishing of $\left.B_{\cal
S}\right\vert_{{\rm large}\,\beta_0}(u=1)$ carries over to the full
theory. It was shown~\cite{Andersen:2005bj,Andersen:2005mj} that so
long as $B_{\cal S}(u)$ does not get particularly large beyond this
region, e.g. near the $u=\frac32$ renormalon position, the Principal
Values Borel sum in Eq.~(\ref{alt_Sud_SL}) is well under control.
Moreover, the resulting spectrum in hadronic variables
(\ref{Pplus_diff}) has approximately the correct physical support
properties, i.e. it smoothly extends into the non-perturbative
region $P^+<\bar{\Lambda}$ and vanishes for $P^+\lsim 0$. This
highly non-trivial property of the Borel--resummed on-shell spectrum
suggests that non-perturbative corrections in this framework are not
large.

Having obtained this result through resummation, it is interesting
to return to the conventional Sudakov resummation framework, in
which Eq.~(\ref{alt_Sud_SL}) is expanded and computed to a given
{\em logarithmic} accuracy, and see why it fails. The comparison
between the results is shown in Fig.~\ref{fig:FLA_vs_DGE}.
Obviously, there is a qualitative difference: the
fixed--logarithmic--accuracy curves are sharply peaked reaching a
Landau singularity near $P^+=0.5$ GeV where they become complex; in
contrast, the DGE result, which is not affected by Landau
singularities, smoothly extends to the non-perturbative regime
$P^+<\bar{\Lambda}\simeq 0.4$ GeV. Aside from the Landau singularity
issue, one observes that results of increasing logarithmic accuracy
represent a badly divergent series, which is dominated by the
leading $u=\frac12$ renormalon. The divergence sets in early (see
Tables 1 and 2 in Ref.~\cite{Gardi:2005mf}) and therefore increasing
the formal accuracy in this naive approach may result in worse
approximations. In contrast, in the DGE approach the renormalon is
regularized and the associated ambiguity cancels.

Eventually, the dependence of the on-shell decay spectrum on the
adopted Principal Value prescription concerns renormalons at
$u=k/2$, with all integer $k\geq 3$. These give rise to ambiguities
scaling as powers of $\left(\Lambda N/(\lambda m_b)\right)^k$. The
corresponding power corrections can be summed up into a
non-perturbative ``shape function'', ${\cal F}\left(\Lambda
N/(\lambda m_b)\right)$, which multiplies the Sudakov factor. This
function has a clear field--theoretic interpretation\footnote{Note
that this is quite different from what became the standard
terminology in the B-decay community, where the term ``shape
function'' is used as synonymous to the quark distribution in the
meson, defined with some ultraviolet cutoff.} as the \emph{moment
space ratio} between the quark distribution in the meson and that in
an on-shell heavy quark~\cite{BDK}. A priori, power corrections on
the soft scale could affect the entire peak region. However, based
on the renormalon ambiguities we actually find the situation is
quite different: the absence of the first two powers and the
inherent suppression of higher powers that is dictated by the
structure of Sudakov exponent (Eqs. (3.50) and (3.51) in
Ref.~\cite{Andersen:2005mj}) suggest that the shape function should
mainly affect high moments $N>\lambda m_b/\Lambda$ and therefore be
important only in the close vicinity of the endpoint,
$P^+\longrightarrow 0$. It amounts to small corrections elsewhere.

Given the properties of the resummed spectrum and the relatively
minor role of non-perturbative corrections in this approach, one can
directly use the perturbative on-shell result as an approximation to
the meson decay spectrum; in this approximation ${\cal
F}\left(\Lambda N/(\lambda m_b)\right)=1$. This was suggested in
Ref.~\cite{Andersen:2005bj} where predictions for the
$\bar{B}\longrightarrow X_s \gamma$ spectrum were obtained. In
addition the first few central moments with a varying lower cut on
the photon energy were computed. Soon after, first results for these
moments were published by the BaBar collaboration, which agree well
with the predictions
--- see Fig.~4 in~\cite{Gardi:2005mf}. A similar comparison with
Belle results~\cite{Belle05} was done later on and is shown in
Figs.~\ref{cut_mom_av} and~\ref{cut_mom_va}.

\begin{figure}[t]
\includegraphics[angle=0,width=19.5pc]{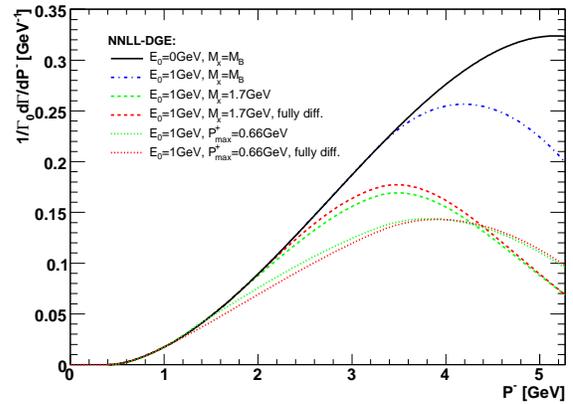}
\caption{The $P^-$ spectrum in $\bar{B}\longrightarrow X_u l
\bar{\nu}$ as calculated by DGE~\cite{Andersen:2005mj}, after
integration over $P^+$ and $E_l$ in four different situations: no
cuts (full), $E_l>1$ GeV cut only (dotdashes), $E_l>1$ GeV  with
$P^+P^-<M_X^2=(1.7\, {\rm GeV})^2$ (dashes), and $E_l>1$ GeV with
$P^+<P^+_{\max}=0.66$~GeV (dots).
\label{fig:diff_Pminus_different_cuts}} \vspace*{-10pt}
\end{figure}

\begin{figure}[t]
\includegraphics[angle=0,width=19.5pc]{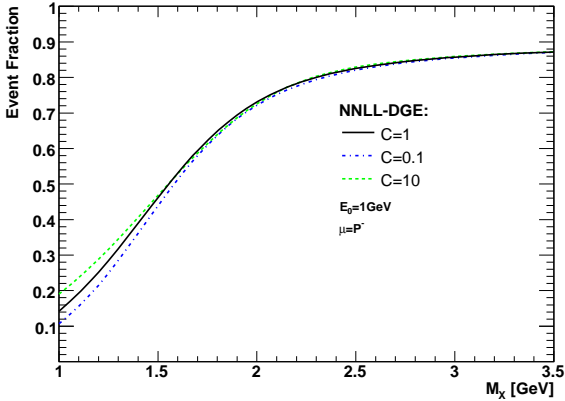}
\caption{The event fraction in $\bar{B}\longrightarrow X_u l
\bar{\nu}$ within the range $P^+P^-<M_X^2$ and $E_l>1$ GeV, as
calculated by DGE, plotted as a function of the cut value $M_X$.
Different assignments for the unknown parameter $C$ (see Sec.~3.2
in~\cite{Andersen:2005mj}), which is proportional to $B_{\cal
S}(u=\frac32)$, are used.\label{fig:Mx_C_dep}} \vspace*{-10pt}
\end{figure}

The good agreement~\cite{Gardi:2005mf} of the on-shell calculation
with the measured  $\bar{B}\longrightarrow X_s \gamma$ decay
spectrum provided a strong incentive to apply this approach to
$\bar{B}\longrightarrow X_u l \bar{\nu}$. This was done in
Ref.~\cite{Andersen:2005mj} where $\vert V_{ub}\vert$ was extracted
for the first time directly based on resummed perturbation theory,
without relying on any parametrization of the spectral shape. As
explained above, the resummation applies to the fully differential
width. This means that the measurement of any partial branching
fraction can be readily translated\footnote{The DGE calculation has
been implemented numerically in \texttt{C++} facilitating
phase--space integration with a variety of cuts. The program is
available at {\tt
http://www.hep.phy.cam.ac.uk/$\sim$andersen/BDK/B2U/}. } into a
measurement of $\vert V_{ub}\vert$.

Fig.~\ref{fig:diff_Pminus_different_cuts} presents the computed
event--fraction as a function of $P^-$. It shows the distribution
for two cuts that have been used by Belle~\cite{Bizjak:2005hn} to
discriminate charm: an upper cut on the hadronic invariant mass at
$M_X=1.7$ GeV, and an upper cut on $P^+$, at $P^+_{\max}=0.66$ GeV.
In both cases a lower cut on the lepton energy $E_l> 1\,{\rm GeV}$
is applied as well. Belle data in these measurements are:
\begin{eqnarray}
&&\hspace*{-20pt}\Delta {\cal B}(P^+P^-<(1.7 \,{\rm GeV})^2) ={
1.24\cdot  10^{-3}} \,\, (\pm  13.4\%)\nonumber \\
\nonumber&&\hspace*{-20pt}\Delta {\cal B}(P^+<0.66\,{\rm GeV}) ={
1.10\cdot 10^{-3}} \,\, (\pm  17.2\%),
\end{eqnarray}
where the numbers in the brackets represent the total experimental
error. The DGE results for the corresponding event fractions are:
\begin{eqnarray}
&&\hspace*{-20pt}{  R_{\rm cut}}(P^+P^-<(1.7\,{\rm GeV})^2) ={
0.615}\,\, (\pm 9.6\%)\nonumber \\ \nonumber&&\hspace*{-20pt}{
R_{\rm cut}}(P^+<0.66\,{\rm GeV})={ 0.535}\,\, (\pm 15.2\%).
\end{eqnarray}
Using
\begin{eqnarray}
\label{Delta_cal_B_th_general} &&\hspace*{-20pt}\Delta {\cal B}({
\bar{B}\longrightarrow X_u l \bar{\nu}}\, \,{\rm { restricted}\,\,
phase\,\, space}) \nonumber \\&&= \tau_B { \Gamma_{\tot}}\left({
\bar{B}\longrightarrow X_u l \bar{\nu}}\right) \times {  R_{\rm
cut}},
\end{eqnarray}
and
$\Gamma_{\tot}\left(\bar{B}\longrightarrow X_u l \bar{\nu}\right)=
{\left\vert V_{ub}\right\vert^2} \times 66.5 \pm 4 \,{\rm ps}^{-1}$,
we obtain~\cite{Andersen:2005mj}
\begin{eqnarray}
\left\vert V_{ub}\right\vert&=&\Big(4.35 \,\pm\, 0.28_{[{\rm
exp}]}\,\pm\,0.26_{[{\rm th}]}\Big)\cdot 10^{-3} \nonumber \\
\left\vert V_{ub}\right\vert&=&\Big(4.39 \,\pm\, 0.36_{[{\rm
exp}]}\,\pm\,0.40_{[{\rm th} ]}\Big) \cdot 10^{-3}, \nonumber
\end{eqnarray}
respectively. The largest source of uncertainty, in both the
calculation of the total width and in that of the event fraction, is
the value of the short--distance mass $m_b^{\MSbar}$. Apart from a
precise mass, to further improve this determination it would be
necessary to have a complete NNLO perturbative calculation to the
fully differential width. Importantly, we find that the sensitivity
to the details of the quark distribution function is small. An
estimate of this source of uncertainty was obtained
in~Ref.~\cite{Andersen:2005mj} by changing $W_{\cal S}(u)$ in Eq.
(\ref{B_JS_QCD}). Fig.~\ref{fig:Mx_C_dep} shows the result as a
function of the $M_X$ cut. Evidently, the effect is small for
experimentally relevant cuts.

Needless to say, the potential of the DGE approach is far from being
exhausted by this perturvative determination of $\left\vert
V_{ub}\right\vert$. In order to quantify the leading power
corrections that constitute ${\cal F}\left(\Lambda N/(\lambda
m_b)\right)$ one would obviously need careful comparison with the
measured spectrum. Both $\bar{B}\longrightarrow X_s \gamma$ and
$\bar{B}\longrightarrow X_u l \bar{\nu}$ decays can be used for this
purpose.

To conclude, we have shown that by properly resumming the
perturbative expansion, the on-shell approximation provides reliable
predictions for inclusive B decay spectra, facilitating a precise
determination of $\left\vert V_{ub}\right\vert$. The calculation of
the Sudakov exponent as a Borel sum guarantees renormalization and
factorization scale invariance, and opens the way for incorporating
valuable information on the large--order behavior of the series. At
the end of the day, the main advantage of this approach stems from
the possibility to use the inherent infrared safety of decay
spectra, which extends beyond logarithmic accuracy.

\end{document}